\documentclass[letterpaper,twoside,twocolumn]{revtex4}
\usepackage[T1]{fontenc}
\usepackage[utf8]{inputenc}
\usepackage{amsmath}
\usepackage{amssymb}
\usepackage{graphicx}
\usepackage{esint}

\makeatletter

\@ifundefined{textcolor}{}
{%
 \definecolor{BLACK}{gray}{0}
 \definecolor{WHITE}{gray}{1}
 \definecolor{RED}{rgb}{1,0,0}
 \definecolor{GREEN}{rgb}{0,1,0}
 \definecolor{BLUE}{rgb}{0,0,1}
 \definecolor{CYAN}{cmyk}{1,0,0,0}
 \definecolor{MAGENTA}{cmyk}{0,1,0,0}
 \definecolor{YELLOW}{cmyk}{0,0,1,0}
}

\makeatother

\usepackage[english]{babel}
\begin{document}

\title{Bell inequalities and entanglement at quantum phase transition in
the XXZ Model}

\author{L. Justino, Thiago R. de Oliveira }

\affiliation{Instituto de Física, Universidade Federal Fluminense, Av. Gal. Milton
Tavares de Souza s/n, Gragoatá, 24210-346, Niterói, RJ, Brazil}
\begin{abstract}
Entanglement and violation of Bell inequalities are aspects of quantum
nonlocality that have been often confused in the past. It is now known
that this equivalence is only true for pure states. Even though almost
all the studies of quantum correlations at quantum phase transitions
only deals with entanglement, we here argue that Bell inequality can
also reveal a general quantum phase transition. This is also shown
for a particular case of two spin-$\frac{1}{2}$ in an infinity one
dimensional chain described by the XXZ model. In this case, the Bell
inequality is able to signal not only the first order phase transition,
but also the infinity order Kosterlitz-Thouless quantum phase transition,
which can not revealed neither by the energy of the system nor by
the bipartite entanglement. We also show that, although the nearest-neighbors
spins are entangled they, unexpectedly, never violate the Bell inequality.
This indicates that the type of entanglement which is relevant for
quantum phase transition is not trivial, \emph{i.e.}, can not be revealed
by the Bell inequality.
\end{abstract}
\maketitle

\section{{\normalsize Introduction}}

The concept of entanglement has evolved since it has first appeared
in 1935, as a ``spooky action at a distance'', \emph{i.e.}, the
possibility of one system to influence another instantaneously, even
at large distances \cite{EPR}. Only in 1964, John Bell moved the
concept from the philosophical debates to the laboratory, using inequalities
- now known as Bell inequalities \cite{Bell}. Entanglement was then
a property of states which could not be described by realistic local
theories. Perhaps the more drastic evolution of the concept came in
1989, with the work of R. F. Werner \cite{Werner}. Werner defined
entangled states as the ones which could not be created using local
operations on the systems with classical communication between the
parts; these operations may create only classical correlations. However,
Werner surprisingly found mixed states which were entangled, according
to his definition, but did not violate the Bell inequality. Entanglement
and non-locality became distinct, but related, concepts.

Even though entanglement has all these facets, the word is usually
used to designate quantum correlations that are ``stronger'' than
classical ones. On the other hand, phase transitions (\emph{e.g.},
magnetic {\small phase)} are generally related to an emerging order
with origin in long range correlations. Quantum phase transitions
(QPT) are the ones occurring at zero temperature, where the systems
is in a pure quantum state. Therefore, we may expect these long range
correlations to come from entanglement. 

Studies trying to link entanglement and QPT first appeared in 2002
for specific models \cite{Osterloh,Osborne}. Later, it was shown
that, in principle, entanglement should inherit the non-analytical
behavior of the ground state energy at any QPT; since both originate
in the ground state \cite{Wu 04,Wu 06}. However, accidental non-analytical
behavior may also appear, or disappear, due to the maximization process
involved in entanglement definitions \cite{MYang}. Nowadays, there
are many studies concerning the link between entaglement and QPT.
Their results can be favorable or unfavorable to it, depending of
the specific model and the measurement of entanglement used (for a
review see \cite{Amico}). Nonetheless, to find such a link, all these
studies used the Werner definition of entanglement and did not analyze
non-local properties of the states. 

In this work, we complement the study of the relation between entanglement
and QPT, analyzing the non-local aspects of quantum phase transition
using Bell inequalities. \emph{The objective is to find: i) if entangled
states, that may play a role in a quantum phase transition, are also
non-local; ii) if the Bell Inequality is able to signal QPT}. We argue
that, in principle, the Bell Inequality should also signal quantum
phase transitions. A particular case of two spins in an infinite one-dimensional
spin-$\frac{1}{2}$ XXZ Heisenberg chain was then studied. The study
of non-local aspects of QPT is very recent, with only one analysis
of a particular case of the one dimensional XY Heisenberg chain with
a transverse magnetic field published\cite{Batle}%
\footnote{After finishing this work we become aware of two related studies \cite{Deng,Altintas}.
For the sake of completeness, we should also mention the study of
Bell Inequality for only two spins interacting via Heisenberg coupling,
and not at a quantum phase transition \cite{Zanardi}. %
}. 

In order to be self contained, we organize the article as follows:
In Sec. II we give a brief introduction about entanglement and non-locality
through Bell Inequality, highlighting its differences. Sec. III describes
the XXZ model and its relevant quantities, in particular how to compute
its correlation functions. The mathematical expression for the entanglement
and Bell Inequalities in the specific case of the XXZ model are given
in Sec. IV. The main results and a detailed discussion of their implications
are in Sec. V and we conclude in Sec. VI.

\section{Entanglement and non-locality}

Bell inequalities investigate correlations between two parties who
share a quantum state. In the case of two spins 1/2 particles, each
side, A and B, choose one direction, $\hat{a}_{i}$ and $\hat{b}_{j}$,
and measures, simultaneously and independently, the observables $\mathbf{A_{i}=}\hat{a}_{i}.\mathbf{\sigma}$
and $\mathbf{B_{j}=}\hat{b}_{j}.\mathbf{\sigma}$; with $\mathbf{\mathbf{\sigma}}=(\sigma^{x},\sigma^{y},\sigma^{z})$.
If each part picks two directions, Bell inequalities state that any
theory which is\emph{ both realistic and local} gives %
\footnote{In this paper we used the Bell-CHSH inequality, so where we see Bell
inequality we read Bell-CHSH inequality. Actually, the inequality
derived by Bell \cite{Bell} was a little different and not suitable
to experimental verification, while the inequality above was written
by Clauser, Horne, Shimony and Holt \cite{CHSH}, hence CHSH inequality,
and is the one usually tested experimentally.%
}

\begin{align}
|\langle B_{CHSH}\rangle| & =|\mathbf{\langle A_{1}}\otimes\mathbf{B_{1}}\rangle+\mathbf{\langle A_{1}}\otimes\mathbf{B_{2}}\rangle+\mathbf{\langle A_{2}}\otimes\mathbf{B_{1}}\rangle\nonumber \\
 & -\mathbf{\langle A_{2}}\otimes\mathbf{B_{2}}\rangle|\leq2.
\end{align}
The value obtained for $\langle B_{CHSH}\rangle$ depends both on
the state and on the directions chosen. So, for a given state one
should maximize over all directions defining a Bell measurement as:

\begin{equation}
{\cal B}=\max_{\{\hat{a}_{i},\hat{b}_{j}\}}|\langle B_{CHSH}\rangle|.\label{eq: Bell Operator}
\end{equation}
The states for which ${\cal B}>2$ are the ones which cannot be described
by a realistic local theory and were then associated with non-classical
correlation or non-locality %
\footnote{Note that, if a given state does not violate the inequality above,
we can only say that this specific combination of correlations can
be described by a realistic local theory. One may still find another
inequality violated by the state. Fortunately, for 2 qubits, and only
2 measurement directions per particle, it is known that the CHSH inequality
is complete: a state satisfying it, will also satisfy any other inequality.
But we can only really state that a state is local if it does not
violate any inequality for any kind of experiment; measures in any
number of directions, for example. Here, we will not be so accurate
and may use the term ``local'' for states which do not violate the
CHSH inequality.%
}. Clearly, states which can be written in a product or separable form,
$|\phi_{A}\rangle|\phi_{B}\rangle$, are local. These states are known
as separable or not entangled.

In 1989, Werner \cite{Werner} proposed a definition of separability
for mixed states from an operational point of view: separable states
are the ones which can be produced by local operations and classical
communication (LOCC). They can be written as 
\begin{equation}
\rho_{AB}=\sum_{k}p_{k}\rho_{A}^{(k)}\otimes\rho_{B}^{(k)},
\end{equation}
with $0\leq p_{k}\leq1$ and $\sum_{k}p_{k}=1$. Separable mixed states
can be correlated, but this correlation is a classical one that comes
from the probabilities $p_{k}$. Surprisingly, in the same work, Werner
constructed states which are entangled by his definition, but nonetheless
do not violate the Bell inequality. Entanglement, although necessary,
is not a sufficient condition for a mixed state to be nonlocal; Bell
inequality does not capture all ``kinds of entanglement''.

For pure bipartite states, only in 1991, Gisin proved that entanglement
is a necessary and sufficient condition for a state to be non-local
\cite{Gisin}. Until nowadays, there are many attempts to draw a unifying
picture between non-locality and entanglement by generalizing the
Bell inequality. Popescu, in 1995, for example, showed that an entangled
state not violating Bell, could be made to violate it after some processing
with LOCC \cite{Popescu}. Another possibility, is to allow each part
to make a measurement in a third direction. This originates a new
inequality that is violated by states which do not violated CHSH inequality
\cite{Gisin-2}. For a introduction and review on similar attempts
see \cite{Liang}. There are also many other definitions of non-locality,
and all of them show some anomaly with entanglement \cite{Scarani}.
For example, while for two qubits pure states the degree of violation
of the Bell inequality is a increasing function of the entanglement
of the state, this is not true for two qudits. Finally, there are
also other definitions of quantum correlation, as the discord which
has also been studied at quantum phase transitions \cite{Sarandy,Werlang,Dillenschneider}.

\section{{\normalsize Spin-$\frac{1}{2}$ XXZ model}}

We now describe the physics of the model we intend to use to study
the relationship between entanglement and non-locality: an one dimensional
spin-$\frac{1}{2}$ chain where the spins interact through anisotropic
Heisenberg interaction. It is known as XXZ model and for N particles
its Hamiltonian is given by 
\begin{equation}
H=\sum_{j=1}^{N}[S_{j}^{x}S_{j+1}^{x}+S_{j}^{y}S_{j+1}^{y}+\Delta S_{j}^{z}S_{j+1}^{z}],
\end{equation}
where $S_{j}^{u}=\sigma_{j}^{u}/2$ ($u=x,y,z$), $\sigma_{j}^{u}$
are the Pauli spin-$\frac{1}{2}$ operators on site $j$, $\Delta$
is the anisotropy parameter, and $\sigma_{j+N}^{u}=\sigma_{j}^{u}$.
The XXZ model cannot be diagonalized, but its energy spectrum can
be obtained by Bethe ansatz. The Hamiltonian has two symmetries: i)
a discrete parity $\mathbb{Z}_{2}$ symmetry over the plane $xy$:
$\sigma^{z}\rightarrow-\sigma^{z}$, and ii) a continuous $U(1)$
symmetry that rotates the spins in the $xz$ plane by any angle $\theta$
. The $\mathbb{Z}_{2}$ symmetry implies that $\langle\sigma_{i}^{z}\rangle=0$
and $\langle\sigma_{i}^{x}\sigma_{j}^{z}\rangle=\langle\sigma_{i}^{y}\sigma_{j}^{z}\rangle=0$,
while the $U(1)$ symmetry implies that $\langle\sigma_{i}^{x}\rangle=\langle\sigma_{i}^{y}\rangle=0$,
$\langle\sigma_{i}^{x}\sigma_{j}^{x}\rangle=\langle\sigma_{i}^{y}\sigma_{j}^{y}\rangle$
and $\langle\sigma_{i}^{x}\sigma_{j}^{y}\rangle=0$ . However, in
the thermodynamical limit a quantum phase transition occur and breaks
the discrete $\mathbb{Z}_{2}$ symmetry; a continuous symmetry may
not be broken at one-dimension even at zero temperature. The model
has three phases: 
\begin{description}
\item [{i)}] $\Delta\leq-1$: the system is in a ferromagnetic phase with
all the spins pointing in the same direction. There is a first order
quantum phase transitions (1QPT) at the critical point $\Delta=-1$
.
\item [{ii)}] $-1<\Delta<1$: the system is in a gapless phase, where the
correlation decays polynomially.
\item [{iii)}] $\Delta>1$: the system is in the anti-ferromagnetic phase.
The critical point at $\Delta=1$ is of infinite order or Kosterlitz-Thouless
quantum phase transitions (KT-QPT).
\end{description}
The Bethe ansatz solution gives the ground state energy \cite{Shiroishi,Yang}:
{\small 
\begin{equation}
e_{0}(\Delta)=\begin{cases}
-\frac{\Delta}{4}, & \Delta\leq-1,\\
\frac{\Delta}{4}+\frac{\sin{\pi\nu}}{2\pi}\int_{-\infty+\frac{i}{2}}^{\infty+\frac{i}{2}}dx\frac{1}{\sinh{x}}\frac{\cosh{\nu x}}{\sinh{\nu x}} & ,-1<\Delta<1\\
\frac{1}{4}-\ln2, & \Delta=1,
\end{cases},
\end{equation}
}where $\Delta=\cos{\pi\nu}$. For $\Delta>1$ we just need to change
$\nu=i\phi$ in (6).

The correlation functions, on the other hand, cannot be easily obtained
from the Bethe ansatz solution. They are given in terms of very complicated
multiple integrals, and solving them is a non trivial mathematical
problem. However, for nearest neighbors we can obtain the correlation
from $e_{0}(\Delta)$:

\begin{equation}
\langle\sigma_{i}^{z}\sigma_{i+1}^{z}\rangle=4\frac{\partial e_{0}(\Delta)}{\partial\Delta},
\end{equation}
 
\begin{equation}
\langle\sigma_{i}^{x}\sigma_{i+1}^{x}\rangle=\langle\sigma_{i}^{y}\sigma_{i+1}^{y}\rangle=\frac{1}{2}(4e_{0}(\Delta)-\Delta\langle\sigma_{i}^{z}\sigma_{i+1}^{z}\rangle).
\end{equation}
For spins further apart, progress has been slow, but there are already
some expressions available up to third neighbors \cite{Shiroishi}.
We will not show them here, since they are too lengthy %
\footnote{There are some typos in equations (19) and (20) from \cite{Shiroishi}.
In (19) we only need to sum a $-\frac{c_{1}}{\pi s_{1}}\zeta_{\nu}$.
In (20) we need to go to \cite{Kato} (note that equation (5.4) has
the same typo) and use equations (5.10), (B.11) and (B.12) to calculate
and find the typo in $\langle\sigma_{i}^{x}\sigma_{i+3}^{x}\rangle$%
}.

We are interested in the entanglement and non-local properties of
two spins in the chain. These can be determined by the reduced density
matrix of the two spins, which can be obtained from the magnetizations
and correlations. Actually, the density matrix of any two spins-$\frac{1}{2}$can
be expressed as {\small 
\begin{equation}
\rho_{i(i+r)}=\frac{1}{4}\left[\mathbb{I\otimes\mathbb{I}}+\mathbf{p.\sigma}\otimes\mathbb{I}+\mathbb{I}\otimes\mathbf{q.\sigma}+\sum_{u,v}t_{r}^{uv}\sigma_{i}^{u}\otimes\sigma_{i+r}^{v}\right],
\end{equation}
with} $r$ being the distance between the sites, $\mathbf{p=\langle\sigma}\otimes\mathbb{I}\rangle$,
${\bf q}=\langle\mathbb{I}\otimes\mathbf{\sigma}\rangle$, and $t_{r}^{uv}=\langle\sigma_{i}^{u}\sigma_{i+r}^{v}\rangle$.
Due to the Hamiltonian symmetries, only $\{t_{r}^{xx},t_{r}^{yy},t_{r}^{zz}\}$
do not vanish and $t_{r}^{yy}=t_{r}^{xx}$. Translation invariance
also implies that the correlation functions for two-sites depend only
on the distance between the sites ($r$), being independent of $i$.
So for the XXZ model we have
\begin{equation}
\rho_{r}=\frac{1}{4}\left(\begin{array}{cccc}
1+t_{r}^{zz} & 0 & 0 & 0\\
0 & 1-t_{r}^{zz} & 2t_{r}^{xx} & 0\\
0 & 2t_{r}^{xx} & 1-t_{r}^{zz} & 0\\
0 & 0 & 0 & 1+t_{r}^{zz}
\end{array}\right).\label{eq: rho XXZ}
\end{equation}

Remark that in the ordered phases, ferro and anti-ferromagnetic, due
to the spontaneous symmetry breaking (SSB), a finite value of the
magnetization will emerge: $\langle\sigma_{i}^{z}\rangle=m$. This
should be taken into account, complicating a little bit the expression
for $\rho$. However, in many studies this is not done, and the simpler
expression, preserving the symmetries, is used. In this case, the
ground state used is not the one which breaks the symmetry and is
stable against external perturbations, but a superposition of the
two degenerated ground states with opposite magnetizations. This could
be justified for the study of very small systems where the SSB does
not happen, but not for real macroscopic systems. It is true, however,
that all the thermodynamical quantities, which comes from the energy
eigenvalues, are not affected by this choice. Nonetheless, entanglement
properties may be affected by this choice, since it also depends on
the eigenvectors \cite{Oliveira,Osterloh06}. Last, besides the symmetric
superposition of the two ground states, one could also think of an
equiproportional mixture of the two, that would emerge if one takes
the limit of temperature going to zero.

\section{{\normalsize Bell measurement and concurrence}}

In general, checking a Bell inequality violation is a high-dimensional
variational problem. We have to optimize over the various possible
measurements settings that each observer performs (see Eq. (\ref{eq: Bell Operator})).
One simplification is that the value of $B_{CHSH}$ and $\mathcal{B}$
does not depend on local properties, the magnetizations $\mathbf{p}$
and $\mathbf{q}$, but only on the correlations $t^{uv}$. For two
spin-$\frac{1}{2}$ particles a closed analytic expression for the
maximum violation of the Bell inequality was obtained by the Horodeckis
\cite{Horodecki} and is given in in terms of the matrix $U=T^{T}T$
by 
\begin{equation}
{\cal B}=2\sqrt{u+u'},
\end{equation}
with $u$ and $u'$ the two largest eigenvalues of $U$ and $T$ the
$3\times3$ matrix build from the correlations $t^{uv}$. As $u$
and $u'$ are combinations of the correlation functions, we can rewrite
it in terms of the correlations.

The expression for ${\cal B}$ in terms of the correlations is very
cumbersome for general states. But, because of the symmetries of the
XXZ model, see Eq. (\ref{eq: rho XXZ}), it becomes quite simple.
In fact, the matrix $T$ is already diagonal, with $t^{xx}$ , $t^{yy}$
and $t^{zz}$ as its diagonal elements and $t^{yy}=t^{xx}$ . Therefore,
the eigenvalues of $U$ are $(t^{xx})^{2}$ and $(t^{zz})^{2}$ and
we only have to check which is greater. The expression becomes 
\begin{equation}
\begin{array}{c}
{\cal B}_{r}=2\max\{\sqrt{2\langle\sigma_{i}^{x}\sigma_{i+r}^{x}\rangle^{2}},\sqrt{\langle\sigma_{i}^{x}\sigma_{i+r}^{x}\rangle^{2}+\langle\sigma_{i}^{z}\sigma_{i+r}^{z}\rangle^{2}}\}.\end{array}\label{eq: Bell XXZ}
\end{equation}
To have non-locality, we need either
\begin{itemize}
\item i)$|\langle\sigma_{i}^{x}\sigma_{i+r}^{x}\rangle|\geq|\langle\sigma_{i}^{z}\sigma_{i+r}^{z}\rangle|\,\text{and}\,|\langle\sigma_{i}^{x}\sigma_{i+r}^{x}\rangle|>\frac{1}{\sqrt{2}}$
\end{itemize}
or
\begin{itemize}
\item ii)$|\langle\sigma_{i}^{x}\sigma_{i+r}^{x}\rangle|<|\langle\sigma_{i}^{z}\sigma_{i+r}^{z}\rangle|\text{and}\langle\sigma_{i}^{x}\sigma_{i+r}^{x}\rangle^{2}+\langle\sigma_{i}^{z}\sigma_{i+r}^{z}\rangle^{2}>1.$
\end{itemize}
To compare the non-locality with entanglement between the two spins,
we will use the concurrence as our entanglement measure; see \cite{Horodecki09}
for a review on entanglement measures. It is a well defined entanglement
measure and is given in terms of the spectrum of $\rho\rho'$, where
$\rho'=\sigma^{y}\otimes\sigma^{y}\rho^{*}\sigma^{y}\otimes\sigma^{y}$
is the time-reversed density matrix. Let $\lambda_{l}$ be the eigenvalues
of $\rho\rho'$ so that $\lambda_{1}\geq\lambda_{2}\geq\lambda_{3}\geq\lambda_{4}$.
Then the concurrence is 
\begin{equation}
{\cal C}=\max\{0,\sqrt{\lambda_{1}}-\sqrt{\lambda_{2}}-\sqrt{\lambda_{3}}-\sqrt{\lambda_{4}}\}.
\end{equation}

The concurrence can also be written in terms of the correlation functions,
but in general the expression is also very cumbersome, since it comes
from solving a polynomial of fourth degree. As in the case of the
Bell inequality, due to the symmetries, for the XXZ model the expression
is simple and given by \cite{Syljuasen} 
\begin{equation}
{\cal C}_{r}=\max\left\{ 0,\frac{2|\langle\sigma_{i}^{x}\sigma_{i+r}^{x}\rangle|-(1+\langle\sigma_{i}^{z}\sigma_{i+r}^{z}\rangle)}{2}\right\} .\label{eq: Concurrence}
\end{equation}

As mentioned before, the Bell inequality is not directly affected
by local properties, the magnetizations, but only depends on the correlations,
given by the matrix $T$. The concurrence, however, may be affected,
since the local magnetizations may appear in its expression. Eq. (\ref{eq: Concurrence})
is only valid when the magnetizations and off-diagonal correlations
are null ($\langle\sigma_{i}^{\alpha}\rangle=0$ and $\langle\sigma_{i}^{x}\sigma_{j}^{z}\rangle=\langle\sigma_{i}^{x}\sigma_{j}^{y}\rangle=\langle\sigma_{i}^{y}\sigma_{j}^{z}\rangle=0$),
and no simple expression can be obtained in terms of the correlation
for the general case. This may have consequences to the effect of
the spontaneous symmetry breaking on these properties. In the case
of the XXZ model, the SSB manifests itself with a finite value of
$\langle\sigma_{i}^{z}\rangle$, which could in principle change the
value of the concurrence, but not of the Bell inequality. However,
as showed in \cite{Syljuasen}, it turns out that for the XXZ model
the concurrence does not change when the SSB is taken into account.

\section{{\normalsize Results}}

\begin{figure}[h]
\centering{}{\small \includegraphics[bb=0bp 0bp 360bp 237bp,clip,scale=0.65]{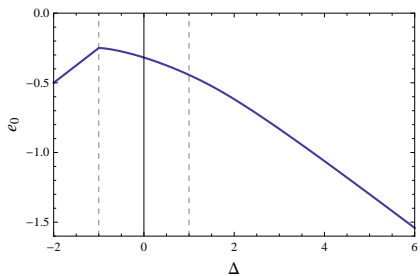}\caption{\label{fig: Energy} Ground state energy per site of the XXZ model
as a function of the anisotropy $\Delta$. There is a first order
quantum phase transition at $\Delta=-1$ from a ferromagnetic to a
gapless phase. At $\Delta=1$ there is an infinite order Kosterlitz-Thouless
phase transition from the gapless to the anti-ferromagnetic phase,
which is not indicated by $e_{0}$.}
}
\end{figure}

\begin{figure}[h]
\begin{centering}
{\small \includegraphics[clip,scale=0.85]{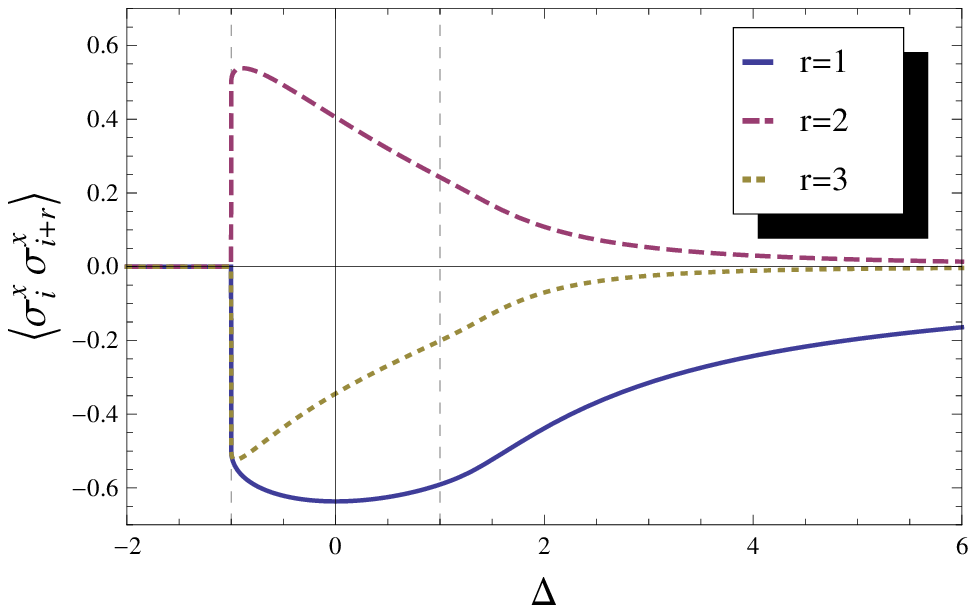}}
\par\end{centering}{\small \par}

{\small \caption{\label{fig: Correlation XX}Correlation function in the $\hat{x}$
direction, $\langle\sigma_{i}^{x}\sigma_{i+r}^{x}\rangle$, for first,
second and third neighbors of the XXZ model as a function of the anisotropy
$\Delta$. It can be seen that the correlation is able to signal the
first order quantum phase transition at $\Delta=-1$, but not the
infinite order quantum phase transition at $\Delta=1$.}
}
\end{figure}

\begin{figure}[h]
\begin{centering}
{\small \includegraphics[clip,scale=0.85]{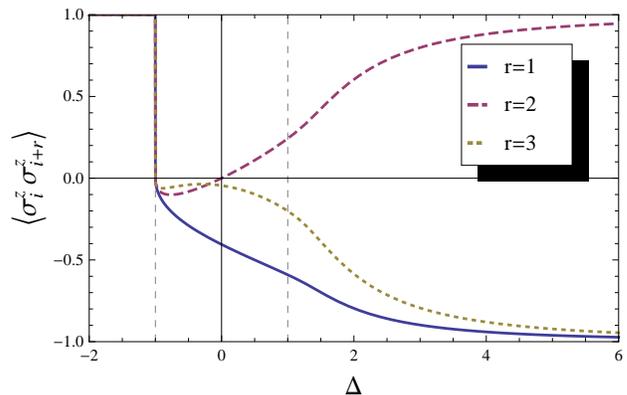}}
\par\end{centering}{\small \par}

\caption{\label{fig: Correlation zz}Correlation function in the $\hat{z}$
direction, $\langle\sigma_{i}^{z}\sigma_{i+r}^{z}\rangle$, for first,
second and third neighbors of the XXZ model as a function of the anisotropy
$\Delta$. It can be seen that the correlation is able to signal the
first order quantum phase transition at $\Delta=-1$, but not the
infinite order quantum phase transition at $\Delta=1$.}
\end{figure}

Before showing the results of non-locality and compare with the entanglement,
let us look at the ground state energy per site $e_{0}$, as shown
in Fig. \ref{fig: Energy}. We can see the non-analytical behavior
of $\frac{\partial e_{0}}{\partial\Delta}$ at the 1QPT $\Delta=-1$
and the nonexistence of non-analyticities at the KT-QPT at $\Delta=1$.
We also show the correlation functions for first, second and third
neighbors in figs. \ref{fig: Correlation XX} and \ref{fig: Correlation zz}.
A non-analytical behavior can also be seen at $\Delta=-1$, and at
least for first neighbors, it is directly related to the one found
in $e_{0}$.

{\small }
{\small \par}

{\small }
{\small \par}

\begin{figure}[h]
\centering{}{\small \includegraphics[clip]{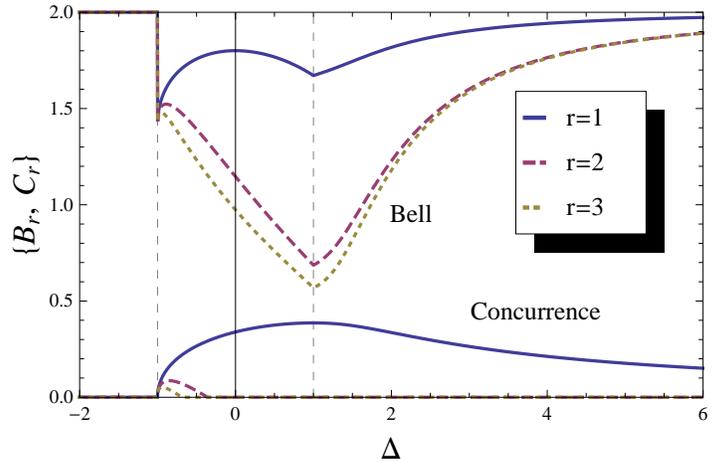}\caption{\label{fig: Bell}Bell measurement (three upper curves) and concurrence
(three lower curves) for first, second and third neighbors of the
XXZ model as a function of the anisotropy $\Delta$. It can be seen
that the Bell inequality is never violated, even when the two spins
are entangled. While both the concurrence and the Bell measurement
are able to signal the first order quantum phase transition at $\Delta=-1$,
only the Bell measurement reveals the infinite order QPT at $\Delta=1$.}
}
\end{figure}

\begin{figure}[h]
\begin{centering}
\includegraphics[clip,scale=0.8]{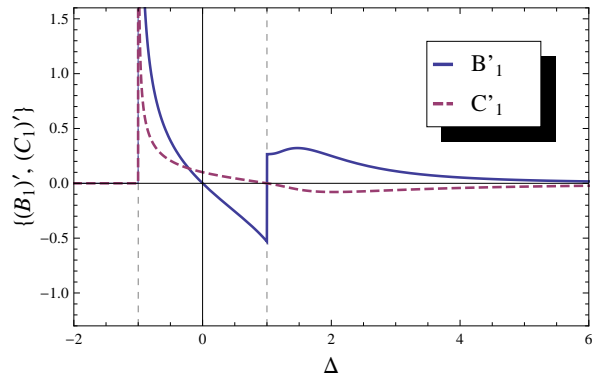}
\par\end{centering}

\caption{\label{fig:Derivative}Derivative of the Bell measurement ${\cal B}_{1}$
and the concurrence $C_{1}$ in relation to $\Delta$ for nearest
neighbors. }

\end{figure}

Last, we review the results of the bipartite entanglement between
two spins, first obtained in \cite{Gu 04}. In Fig. \ref{fig: Bell}
we observe that the entanglement of first neighbors suddenly appears
as they enter the gapless phase, achieve the maximum at the KT-QPT,
and slowly decrease in anti-ferromagnetic phase. Second and third
neighbors are only slight entangled and in a small region on the right
of the 1QPT; the curves can barely be seen in Fig. \ref{fig: Bell}.
The first order phase transition is indicated by the non-analytical
behavior of the concurrence of all the first three neighbors: its
first derivative diverges (see Fig. \ref{fig:Derivative} for the
case of first neighbors)

We now present our results. First we argue that the Bell Inequality
may be as good as entanglement to signal the QPT. The argument is
equivalent and follows directly from the one used in \cite{Wu 04}
for bipartite entanglement and in \cite{Oliveira 07} for multipartite:
for Hamiltonians with only nearest-neighbors interactions not only
the entanglement between nearest-neighbors spins, but also $e_{0}$
can be obtained from the reduced density matrix of the two spins;
they are functions of $\rho_{i,i+1}$. So any non-analytical behavior
in $e_{0}=f(\rho_{i,i+1})$ will also appear in the entanglement $E=g(\rho_{i,i+1})$,
unless the the specific form of the functions $f$ and $g$ cause
the non-analitycal behavior or originate accidental cancellations
(see \cite{MYang} for example). Note that while $f$ is a linear
function of $\rho_{i,i+1},$ $g$ involves the absolute value of $\rho_{i,i+1}$
and maximizations or minimizations procedures.

But such analyze is also trivially true for any property of the two
spins, since they are completely determined by $\rho_{i,i+1}$, including
the Bell measure. \emph{Therefore $\mathcal{B}$ is, in principle,
as good as any entanglement measure to signal a QPT}. But as in the
case of entanglement accidental cancellations may also happen due
to the complicated form $\mathcal{B}$ depends on $\rho_{i,i+1}$:
it also involves the absolute value of $\rho_{i,i+1}$ and a maximization.
Note that signaling the QPT is not the only interest when studying
the behavior of entanglement or non-locality at QPT. One could also
learn about the nature of the correlations involved in the QPT, as
we will argue in the following when looking at the specific case of
the XXZ model.

Given the general argument, we move to our results for the case of
the XXZ model. The value of the Bell measurement, using Eq. (\ref{eq: Bell XXZ}),
is shown in the upper curves of Fig. \ref{fig: Bell} for first (solid
blue curve), second (dashed red curve) and third (doted yellow curve)
neighbors. We observe that none of them violate the Bell inequality,
despite being entangled. But lets first concentrate on the analytical
properties of $\mathcal{B}$. As it may be expected from the arguments
above, it is able to indicate the QPTs. Furthermore, \emph{not only
the first order QPT is revealed by $\mathcal{B}$, but also the Kosterlitz-Thouless
QPT, which is neither indicated by} \emph{the energy nor by the bipartite
entanglement given in terms of the concurrence}. This can also be
seen in Fig. \ref{fig:Derivative} where we plot the derivatives of
the Bell measurement ${\cal B}_{1}$ and the concurrence $C_{1}$
for nearest neighbors. It shows that while the derivative of both
diverges at the first order quantum phase transition ($\Delta=-1$),
only the derivative of the Bell measurement signal the infinite order
quantum phase transition ($\Delta=1$).

It is interesting to note that the success of $\mathcal{B}$ in revealing
the KT-QPT comes from the maximization process, which is usually associated
with negative results in the ability of entanglement to signal a QPT.
More specifically, analyzing Eq. (\ref{eq: Bell XXZ}), we find out
that while in the ferro and anti-ferromagnetic phase $|\langle\sigma_{i}^{x}\sigma_{i+r}^{x}\rangle|\leq|\langle\sigma_{i}^{z}\sigma_{i+r}^{z}\rangle|$
and $\mathcal{B}_{r}=2\sqrt{\langle\sigma_{i}^{x}\sigma_{i+r}^{x}\rangle^{2}+\langle\sigma_{i}^{z}\sigma_{i+r}^{z}\rangle^{2}}$,
in the gapless phase $|\langle\sigma_{i}^{x}\sigma_{i+r}^{x}\rangle|>|\langle\sigma_{i}^{z}\sigma_{i+r}^{z}\rangle|$
and $\mathcal{B}_{r}=2\sqrt{2\langle\sigma_{i}^{x}\sigma_{i+r}^{x}\rangle^{2}}$
. On the other hand, the non-analytical behavior of $\mathcal{B}$
at the first order QPT does originate in $\rho_{i,i+r}$ or in $e_{0}$. 

Just to contrast, we should mention that the non-analytical behavior
of the concurrence does not come from the non-analyticities in $e_{0}$,
but from the maximization involved in the concurrence. Note also that
at the KT-QPT the first neighbor concurrence is maximum, while all
the next neighbors are null almost in the whole phase. However, even
though one can show that, in general, the concurrence may inherit
the non-analytical behavior of $e_{0}$ , it is not possible to show
that it should be maximum at the critical point. Therefore, such a
behavior, should not be an indication of a QPT and may be specific
to the XXZ model; in fact it has been interpreted as a consequence
of the specific symmetries of the model (see Sec. IV-A-2 of \cite{Amico}).
It would be interest to know if the behavior of $\mathcal{B}$ at
the KT-QPT is also only specific to this model.

Last, let us comment on the non-violation of the Bell inequality for
nearest-neighbors, our third result. It is unexpectedly, since these
spins are entangled, and not only slightly. They have a kind of hidden
entanglement, similar to the one present in the Werner states, that
is not revealed by the Bell inequality, being in this sense local.
That entanglement may be relevant for quantum phase transitions has
been suggested by many works \cite{Amico}. \emph{Our results complements
such works,} \emph{indicating that the kind of entanglement that is
relevant for QPT is not trivial: it cannot be revealed by the Bell
inequality}. Note that, one may argue that this should be expected
since we are studying two particles in a huge chain in a pure highly
entangled state. Thus, these two particles are in very mixed states
close to the identity and should not violate the Bell inequality.
But this argument also applies to entanglement: the two particles
should not be entangled. In our case they are, and not slightly. In
fact, despite even slight entangled states may violate Bell, the number
of states violating Bell and the degree of violation increase with
the amount of the entanglement \cite{Verstraete}. We also observe
that second and third neighbors do not violate Bell, but these are
only weakly entangled in a tiny region of the phase diagram. Note
that, even tough our states have a behavior similar to Werner states,
they are not Werner states. Only at $\Delta=1$ can our state be written
in the form of a Werner state. Thus, these are a new class of entangled
states, which do not violate the Bell Inequality.

{\small }
{\small \par}

\begin{figure}[h]
\begin{centering}
\includegraphics[clip,scale=0.85]{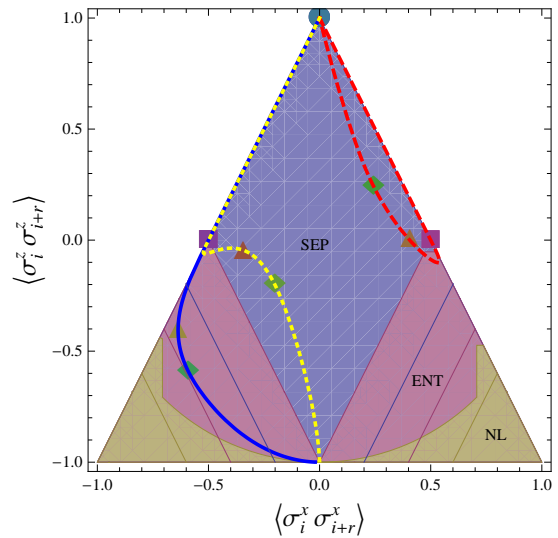}
\par\end{centering}

\caption{\label{fig:Parameter-space}Parameter space of the reduced density
matrix of two spins with the symmetries of the XXZ Hamiltonian. The
greater triangle defines the physical states. The two smaller triangles,
inside and on the lower vertices of the greater, are the entangled
states (ENT). The two smallest regions, also on the lower vertices
of the greatest triangle, are the states violating Bell inequality
(NL). The curves are the ``trajectory'' of the ground state of the
XXZ Hamiltonian as the parameter $\Delta$ is varied for $\rho_{i,i+1}$
(solid blue curve), $\rho_{i,i+2}$ (dashed red curve) and $\rho_{i,i+3}$
(dotted yellow curve). For all the curves we marked fours points that
corresponds to $\Delta=-1$ (full circle), -0.999 (square), 0 (triangle)
and 1 (diamond).}

\end{figure}

Our results can be visualized geometrically in Fig. \ref{fig:Parameter-space},
where we have the parameter space of density matrices of two spin-$\frac{1}{2}$
particles with the symmetries of the XXZ Hamiltonian. Remember $\rho_{i,j}$
only depends on $\langle\sigma_{i}^{x}\sigma_{j}^{x}\rangle$ and
$\langle\sigma_{i}^{z}\sigma_{j}^{z}\rangle$; Eq. (\ref{eq: rho XXZ}).
The greater triangle defines the physical states and the inner diamond
the separable states (SEP). The two smaller triangles, inside and
on the lower vertices of the greater, are the entangled states (ENT).
The two smallest regions, also on the lower vertices of the greatest
triangle, are the states violating Bell inequality (NL). The parallel
lines are contour lines of the concurrence. We can see that it increases
in the direction of the lower vertices of the greatest triangle. The
contour lines of the Bell inequality are parallel to the lines defining
the border of the region NL, but not show. We also plotted the ``trajectory''
of $\rho_{i,i+r}$ for $r=1,2\text{\,\ and\,}3$ as the parameter
$\Delta$ is varied. For all the curves we marked fours points that
corresponds to $\Delta=-1$ (full circle), -0.999 (square), 0 (triangle)
and 1 (diamond). The whole ferromagnetic phase ($\Delta\leq-1$) is
represented by the full circle. As we move from the ferromagnetic
phase in the direction of the gapless ($-1<\Delta<1$) there is a
jump at the point $\Delta=-1$. The state of $\rho_{i,i+1}$ (solid
blue curve) and $\rho_{i,i+3}$(dotted yellow) jumps through the left
side, entering the entangled region, while $\rho_{i,i+2}$ (red dashed
curve) jump to the entangled region through the right side. We can
also observe that $\rho_{i,i+2}$ and $\rho_{i,i+3}$ barely enter
the entangled region and go back to the separable one, while $\rho_{i,i+1}$
goes deeper in the entangled region and stays there in the whole gapless
phase. The maximum value of the concurrence for $\rho_{i,i+1}$, at
the critical point $\Delta=1$, can be seen as the point where the
trajectory tangencies a contour line of the concurrence. At this point
the trajectory also touches a minimum of the Bell measurement, which
happens due to the optimization procedure involved in its definition.

\section{{\normalsize Conclusion}}

We have studied non-locality using Bell inequality at quantum phase
transitions. We first showed that \emph{the Bell Inequality is, in
principle, as good as any entanglement measurement to signal a }quantum
phase transistion. Then, we studied the particular case of two spins
in an one-dimensional spin-$\frac{1}{2}$ XXZ infinite chain at zero
temperature and found that: i) \emph{the Bell inequality is not only
able to reveal the first order quantum phase transition, but also
the Kosterlitz-Thouless quantum phase transition. This can not be
indicated by the energy} \emph{of the system nor by the bipartite
entanglement given in terms of the concurrence.} ii) \emph{the two
spins, despite being entangled, never violate the Bell inequalities:
their correlations can be then described by a realistic local theory.}

Our results suggest that the Bell inequality may contribute to the
study of quantum phase transitions, since they are able to signal
quantum phase transitions - and even the\emph{ }Kosterlitz-Thouless
quantum phase transition, in our specific model. Furthermore, they
may increase our understanding of the nature of the quantum correlations
involved in quantum phase transitions. Again, in our specific model,
it was shown that the type of entanglement which is relevant at the
quantum phase transition is non-trivial, since it cannot be revealed
by the Bell inequality. Therefore, Bell inequality constitute a different
resource, other than entanglement, and could be an alternative and
complementary way to characterize a quantum phase transition, maybe
even infinite order ones.

Our work gives information about the type of quantum correlation involved
at the quantum phase transition of the XXZ model. To better characterize
these quantum correlations, it would be interesting to study other
models and more general Bell inequalities involving more measurement
directions. The study of multipartite Bell inequalities is also of
great interest, despite its technical difficulties. First, obtaining
correlations among three or more spins in condensed matter systems
suffering quantum phase transitions is not trivial. Second, there
is no unique multipartite inequality, but many to be studied. However,
a work on finite chain up to six spins was recently published\cite{Campbell 10}.
\begin{acknowledgments}
Thiago R. de Oliveira thanks some fruitful discussion with Gustavo
Rigolin. Both authors thanks Gustavo Rigolin, Sergio Sanchez and Ernesto
Galvão for careful reading of the manuscript and acknowledge financial
support from the Brazilian agencies CNPq and CAPES-REUNI. This work
was performed as part of the Brazilian National Institute of Science
and Technology for Quantum Information.It would be interesting \end{acknowledgments}

\end{document}